\pdfoutput=1

\documentclass[]{spie}  

\usepackage{amsmath,amsfonts,amssymb}
\usepackage{graphicx}
\usepackage{pdfpages}
\usepackage{siunitx}

\DeclareGraphicsExtensions{.png,.pdf}

\usepackage[colorlinks=true, allcolors=blue]{hyperref}

\title{MUSCAT focal plane verification}

\author[a]{M. Tapia}
\author[b]{P. A. R. Ade}
\author[c,d]{P. S. Barry}
\author[b]{T. L. R. Brien}
\author[e]{E. Castillo-Domínguez}
\author[a]{D. Ferrusca}
\author[a]{V. Gómez-Rivera}
\author[b]{P. Hargrave}
\author[a]{J. L. Hérnandez-Rebollar}
\author[b]{A. Hornsby}
\author[a]{D. Hughes}
\author[a]{J. M. Jáuregui-García}
\author[e]{P. Mauskopf}
\author[a]{D. Murias}
\author[b]{A. Papageorgiou}
\author[g]{E. Pascale}
\author[a]{A. Pérez}
\author[b]{S. Rowe}
\author[a]{C. Tucker}
\author[a]{M. Velázquez}
\author[a]{S. Ventura}
\author[b]{S. Doyle}

\affil[a]{Instituto Nacional de Astrofísica, Óptica y Electrónica, Luis Enrique Erro 1, Santa María Tonantzintla 72840 Puebla, Mexico}
\affil[b]{School of Physics \& Astronomy, Cardiff University, The Parade CF24 3AA, Cardiff, United Kingdom}
\affil[c]{University of Chicago, 5640 S. Ellis Ave., Chicago IL, USA}
\affil[d]{{Argonne National Laboratory, 9700 S. Cass Ave., Lemont IL, USA}}
\affil[e]{SRON-Netherlands Institute for Space Research, Landleven 12, 9747 AD Groningen, Netherlands}
\affil[f]{Arizona State University, Tempe, Arizona, United States of America}
\affil[g]{Dipartimiento di Fisica, La Sapienza Universitá di Roma, Piazzale Aldo Moro 5, 00185 Roma, Italy}

\authorinfo{Further author information: (Send correspondence to M. Tapia)\\E-mail: mbecerrilt@inaoep.mx}

\begin{document} 
\maketitle

\abstract{The Mexico-UK Submillimetre Camera for Astronomy (MUSCAT) is the second-generation large-format continuum camera operating in the 1.1 mm band to be installed on the 50-m diameter Large Millimeter Telescope (LMT) in Mexico. The focal plane of the instrument is made up of 1458 horn coupled lumped-element kinetic inductance detectors (LEKID) divided equally into six channels deposited on three silicon wafers.
Here we present the preliminary results of the complete characterisation in the laboratory of the MUSCAT focal plane. Through the instrument's readout system, we perform frequency sweeps of the array to identify the resonance frequencies, and continuous timestream acquisitions to measure and characterise the intrinsic noise and 1/f knee of the detectors. Subsequently, with a re-imaging lens and a black body point source, the beams of every detector are mapped, obtaining a mean FWHM size of $\sim$3.27 mm, close to the expected 3.1 mm. Then, by varying the intensity of a beam filling black body source, we measure the responsivity and noise power spectral density (PSD) for each detector under an optical load of 300 K, obtaining the noise equivalent power (NEP), with which we verify that the majority of the detectors are photon noise limited. Finally, using a Fourier Transform Spectrometer (FTS), we measure the spectral response of the instrument, which indicate a bandwidth of 1.0--1.2 mm centred on 1.1 mm, as expected.}


\section{Introduction}
\label{sec:intro}

The important advances in recent years in the field of sub-mm/mm astronomy are mostly due to the development of telescopes with higher angular resolution: JCMT-15 m, IRAM-30 m or the ALMA interferometer; and to the introduction of improved detector technology: transition-edge sensors (TES) and kinetic inductance detectors (KID), with the ability to achieve background limited sensitivities in large format focal plane arrays.

In particular, KID technology, which has shown photon-noise-limited performance in this wavelength range,\cite{Yates2011,Mauskopf2014} has achieved notoriety due to the intrinsic ability to couple a large number of detectors to a single microwave feedline for readout through a single pair of coaxial cables. Despite its relatively recent invention,\cite{Day2003} great efforts have been made for its consolidation and on-sky validation in the continuum with NIKA,\cite{Monfardini2010} BLAST-TNG,\cite{Galitzi2014} and OLIMPO; \cite{Paiella2020} and in spectrometry with SuperSpec,\cite{Kovacs2012} and DESHIMA\cite{takekoshi2020}; all with promising results.

It is in this context that the Mexico-UK Submillimetre Camera for Astronomy (MUSCAT) is conceived. MUSCAT is a large-format continuum camera for the Large Millimeter Telescope (LMT) Alfonso Serrano which has a primary-mirror diameter of 50 meters and is located at the summit of the Sierra Negra volcano (4600~m) in Mexico. MUSCAT will operate in the 1.1~mm atmospheric band with 1458 lumped-elements kinetic inductance detectors (LEKID). The combination of the instrument's large pixel count and the LMT's extensive collection area, as well as the site's excellent atmospheric conditions,\cite{Zeballos2016} will allow high-speed mapping observations, providing large-area survey capabilities for a wide range of astronomical purposes\cite{Castillo2018, Tom2018}. 

In this work, we present the results of the complete laboratory characterization of the MUSCAT focal plane. In Section 2, we describe the main characteristics of the detectors, their installation in the focal plane and the set of experiments carried out to quantify their main parameters, including measurements of the transmission $S_{21}$, the beam shape, responsivity, noise equivalent power (NEP), and spectral response. In Section 3, we present the main results of each experiment and in Section 4, we summarize the main conclusions of this work.

\section{Measurements description}
\label{sec::meas:description}

\subsection{Detector arrays description}
\label{sec::array_description}

The morphology of a MUSCAT KID consists of an inductive meander as a photosensitive region, in series with an interdigital capacitor (IDC), and additional capacitive coupling to the readout transmission line (Figure 1a). A continuous aluminum meander forms the inductor which is $3~\si{\micro\metre}$ wide and $25~\si{\nano\metre}$ thick, arranged in a pattern optimised for coupling to both polarizations, covering a square area of $2\times2~\si{\milli\metre}^{2}$. The characteristic resonance frequency of each detector is assigned by adjusting the number and length of the IDC fingers, while the coupling quality factor is maintained by adjusting the length of the coupling capacitor.

\begin{figure}
    \centering
    \includegraphics[width=0.82\textwidth]{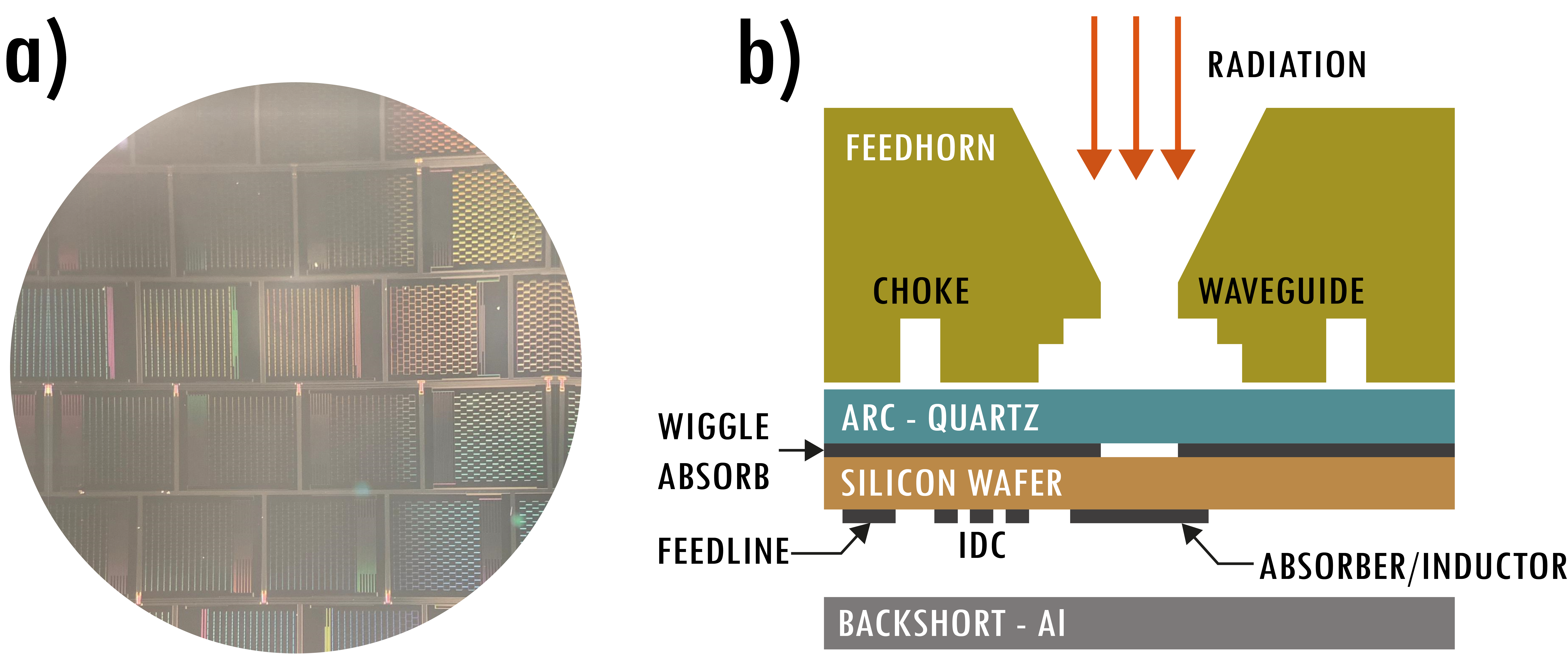}
    \caption{MUSCAT detectors. a) A photograph of LEKID pixels from one of the MUSCAT arrays. b) The configuration of the MUSCAT detectors. Radiation is coupled through an aluminum feed horn with a cylindrical waveguide and choke. A quartz anti-reflection layer helps to impedance match the free space input to the silicon substrate through which the detectors are illuminated. An aluminum mesh layer (referred to as the 'wiggle absorber') is included between the substrate and the quartz to minimise stray light leakage between pixels.}
    \label{fig:ExampleFSM}
\end{figure}

The detector is back-illuminated according to the structure shown in Figure 1b. A feedhorn/waveguide/choke system couples the radiation to the detector. An absorbing layer of aluminium mesh pattern is deposited on the opposite face to the meander and IDC pattern, covered by a $\lambda/4$ anti-reflection layer (AR) of quartz. At the base of the structure, there is an aluminium plate (back-short) placed at a distance tuned to match the impedance of the meander on silicon structure. This design reduces unwanted internal reflections that contaminate surrounding detectors and maximises optical coupling to incoming radiation. The dimensions and distances between these elements were optimised through simulations with Ansys's HFSS suite. A more detailed description of this configuration design can be found in G\'{o}mez-Rivera et al.\cite{Gomez2020}

The MUSCAT detectors are distributed across three Si wafers in a total of six microwave readout channels of 243 detectors each. The detectors in a single readout channel are hex-packed with 1 $F \lambda = 3.1$ mm spacing in nine rows with 27 pixels each, with two channels deposited side by side on each silicon wafer. The resonance frequencies on each channel occupy a bandwidth of $\sim$500~MHz, with an average resonator spacing of $\sim2~\si{\mega\hertz}$. The design coupling quality factor for all resonators with the feedline is $Q_{\mathrm{c}}\sim50000$, to minimise the probability of resonator clashes caused by unwanted shifts of resonances by parasitic capacitive loads, and manufacturing non-uniformities.

For each readout channel, the resonance frequencies are divided into two bands: a low-frequency (LF) band with a total of 135 detectors spaced by 1.4 to 2~MHz; and a high frequency (HF) band with 108 detectors spaced by 1.8--2.4~MHz. There is a 20~MHz gap between them, enough to operate the readout local oscillator without interfering with the resonators.

The detector arrays were manufactured at the University of Chicago using direct-write lithography and aluminium deposition by e-beam evaporation. In each manufacturing process, two channels are deposited on a single wafer, plus four small test arrays with 20 detectors each for characterisation and validation of the pixel design.\cite{Gomez2020}. The three fabricated wafers used are indexed as follows, channels 1-2 correspond to fabrication run MD-M, 3-4 to MD-J, and 5-6 to MD-F. The detectors in MD-F arrays have resonance frequencies that cover the bandwidth of 580--1060 MHz (480 MHz). In contrast, the MD-J and MD-M arrays are identical, and their detectors have a bandwidth of 507--980~MHz (473~MHz). The bandwidth shift at lower frequencies for these latter arrays was added on a second iteration of the design in order to compensate for additional frequency dispersion from the added stray light absorber layer.

The six arrays are mounted on the MUSCAT focal plane as shown in Figure 2, where they operate continuously at 130~mK through a novel closed-cycle cooling system described in Brien et al.\cite{Brien2018} Each of the six channels is readout by a pair of coaxial cables that are heat sunk at each temperature stage of the system (50~K, 4~K and 450~mK) before connecting to the array modules. In addition, DC blocks add further thermal isolation between the centre line of each cable between the 4~K and 450~mK stages. RF attenuators of 20~dB and 10~dB are placed on the input lines at the 4~K and 450~mK stages respectively to reduce thermal noise in the probe signal and a low noise cryogenic amplifier (LNA) is placed on the output lines at 4~K. On each array, a set of spring-loaded pins are used to provide support to the wafer, reducing the mechanical vibrations in the detectors due to the pulse tube cooler (PTC).\cite{Brien2020}

\begin{figure}
    \centering
    \includegraphics[width=1.0\textwidth]{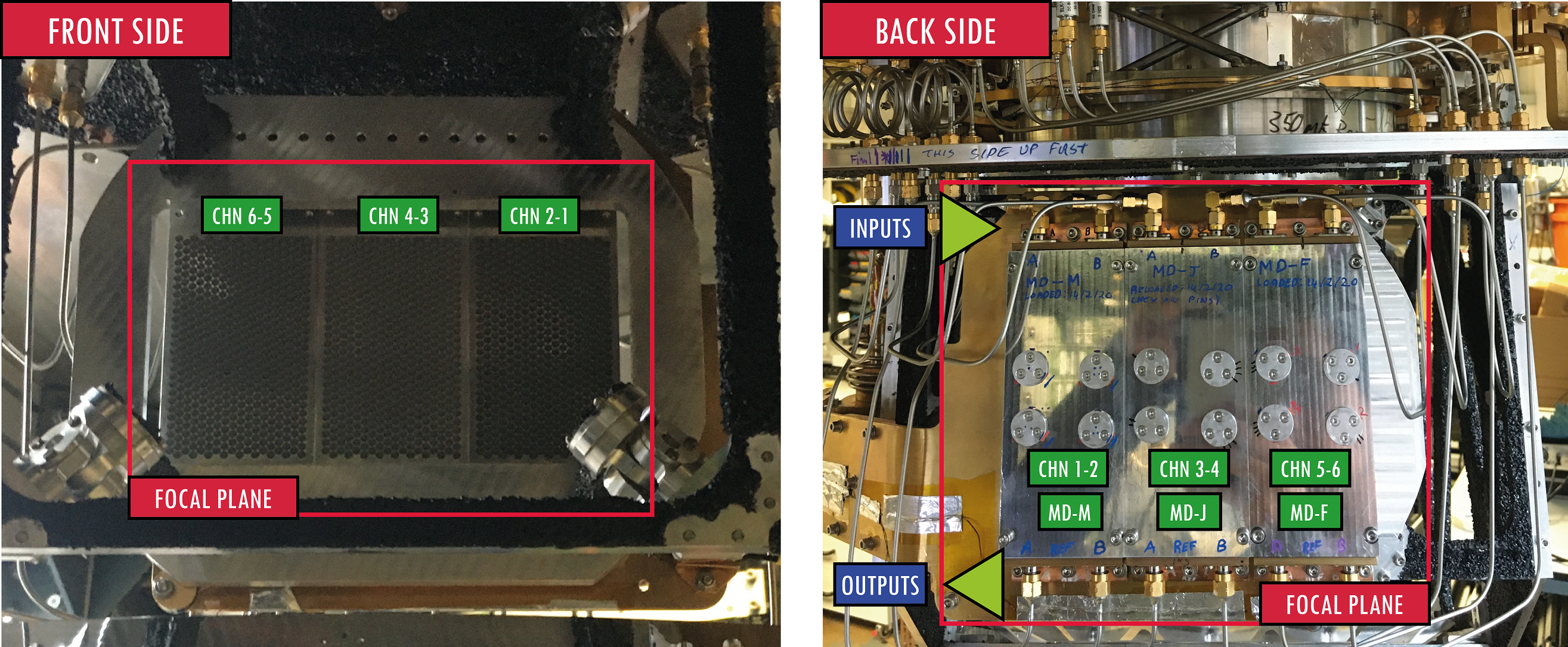}
    \caption{MUSCAT focal plane. Front view (left): a horn block is placed on each pair of detector arrays. Back view (right): from this side, the MD-M, MD-J and MD-F deployment encapsulated detector arrays (from left to right) are installed, connected to the inputs/outputs of channels 1 to 6 (left to right).}
    \label{fig:VNA_meas}
\end{figure}

\subsection{$S_{21}$ transmission and timestreams measurements}

The set of measurements for the characterization of the detectors on the MUSCAT focal plane were carried out with the camera mounted on an aluminum construction frame, suspended at a height of $\sim1.6~\si{\metre}$, with the entrance aperture and filter stack pointing upward, as shown in Figure 3. This frame facilitates the  assembly/disassembly of the cryostat, installation and testing of cold electronics, cooling stages, and the mounting of detectors onto the focal plane, but it is not required for observations at the LMT.

\begin{figure}
    \centering
    \includegraphics[width=0.65\textwidth]{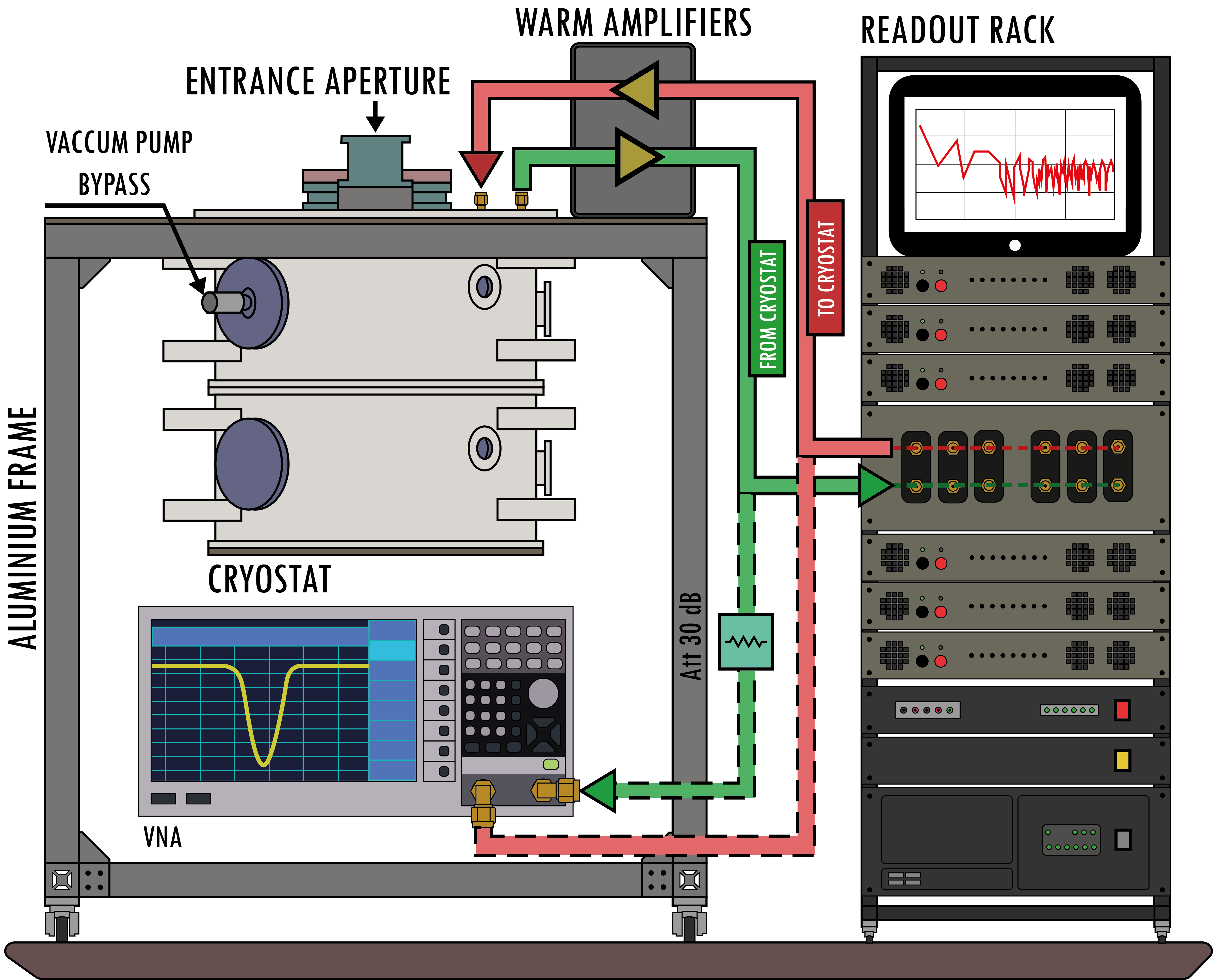}
    \caption{Experimental configuration to measure the transmission and timestreams of detectors in the MUSCAT focal plane. A wide frequency sweep of each of the arrays is performed with a VNA, from which the resonance frequencies of each available detector are extracted. These frequencies are used to generate a tones list for the readout electronics which is then updated with the suitable readout power of each detector. Continuous acquisition of the variation in complex transmission of each tone is then made using six multiplexed readout electronics sets to generating readout timestreams for each detector used to monitor detector response for sensitivity, beam shape and spectral characterisation.}
    \label{fig:VNA_meas}
\end{figure}

First, the cryostat entrance window is covered with an Eccosorb foam sheet, providing a stable 300~K optical load, and the transmission parameters $S_{21}$ of the six channels are measured with the Vector Network Analyzer (VNA). To regulate the power input to the detectors, we place a 30 dB attenuator between port 1 of the VNA and the input to the cryostat. A 30 dB amplifier is then added on the output to increase the signal strength at port 2 of the VNA. For the six channels, we perform a wide frequency sweep between 0.1-2 GHz with a coarse resolution of 1 kHz, to identify the resonators in each array. Then, around each resonator, we repeat the sweeps with enough resolution to extract the physical parameters such resonance frequency, quality factor and microwave power handling, based on the theoretical resonator model from Swenson et al.\cite{Swenson2013} From this information, a tone list is built and the frequency of the local oscillator is set.

Next, we directly connect the six inputs/outputs of the cryostat to the readout system, which is based on a slightly modified version of the BLAST-TNG KID readout system.\cite{Gordon2016} The readout initialization routines are executed, including the setting the hardware configuration of ROACH-2 boards, synthesizers and variable attenuators; setting the local oscillator; loading the probe tones; and, performing the frequency sweeps. The aim is to drive the resonators with as much RF power as possible, to reduce the noise contribution from the first stage cryogenic amplifier, without entering the bifurcation or over-driven regime. The placement of the tones is according to the criteria of the maximum speed of the resonator $I$-$Q$ circles (i.e. the magnitude of the rate of change of the quadrature signals $I$ and $Q$ with respect to the frequency change is maximum).

In this way, the system is prepared for the continuous and parallel acquisition of the quadrature $I$ and $Q$ signals modulated by the individual response of each resonator. From these timestreams, together with parameters taken from the respective frequency sweeps, the resonance frequency shifts $df$, or fractional frequency shifts can be estimated.

\subsection{Beam mapping}

To allow verification of the instrument's point spread function (PSF) in the lab environment without the primary optics of the LMT, a re-imaging lens made from ultra-high-molecular-weight polyethylene (UHMWPE) is placed directly over MUSCAT's window. The beam mapping is performed by raster scanning a blackbody point source over the new image plane with a steps of 1 mm in both axes, this is enough to accurately determine the beam shape of each of the detectors (design FWHM $\sim 3.1~\si{\milli\metre}$). Simultaneously, the timestreams of all the resonators are acquired in parallel, with which we will later generate the position-intensity maps of each detector (see Figure 4).

The blackbody source is a 1500~K Interspectrum, type-2, thermal IR source, mounted behind a chopper wheel in an enclosure with an adjustable aperture of $\sim 2.7~\si{\milli\metre}$ (acting as a point source), which is installed on a mobile Cartesian platform that moves linearly in the X-Y plane. This platform is placed at the new focus of the camera, which for practical reasons, is folded $90\si{\degree}$ towards one side of the frame using a flat mirror.

In addition to revealing the detailed structure of the detector PSFs, the beam map results also provide the correspondence between the detector resonance frequency and position on the focal plane. Small variations in each detector's coupling elements (e.g. horn position relative to the detector) influence the geometrical inductance across the array, creating a shift in $f_{0}$ from its design value. This shift is not the same for all detectors, with resonators in the high-frequency band being more susceptible to this effect, causing dispersion and widening of the array bandwidth. In addition to this, a number of other factors contribute to the dispersion of the resonators. Including, bowing of both the anti-reflection and substrate wafers, variations in the widths and thicknesses of the traces delineating the detectors, scratches, open or short circuits, and variations in thickness of the oxide layer that forms on the surface of aluminium. Furthermore, the total number of measurable resonators is reduced by unacceptable levels of cross-talk between some detectors. 

\begin{figure}
    \centering
    \includegraphics[width=0.75\textwidth]{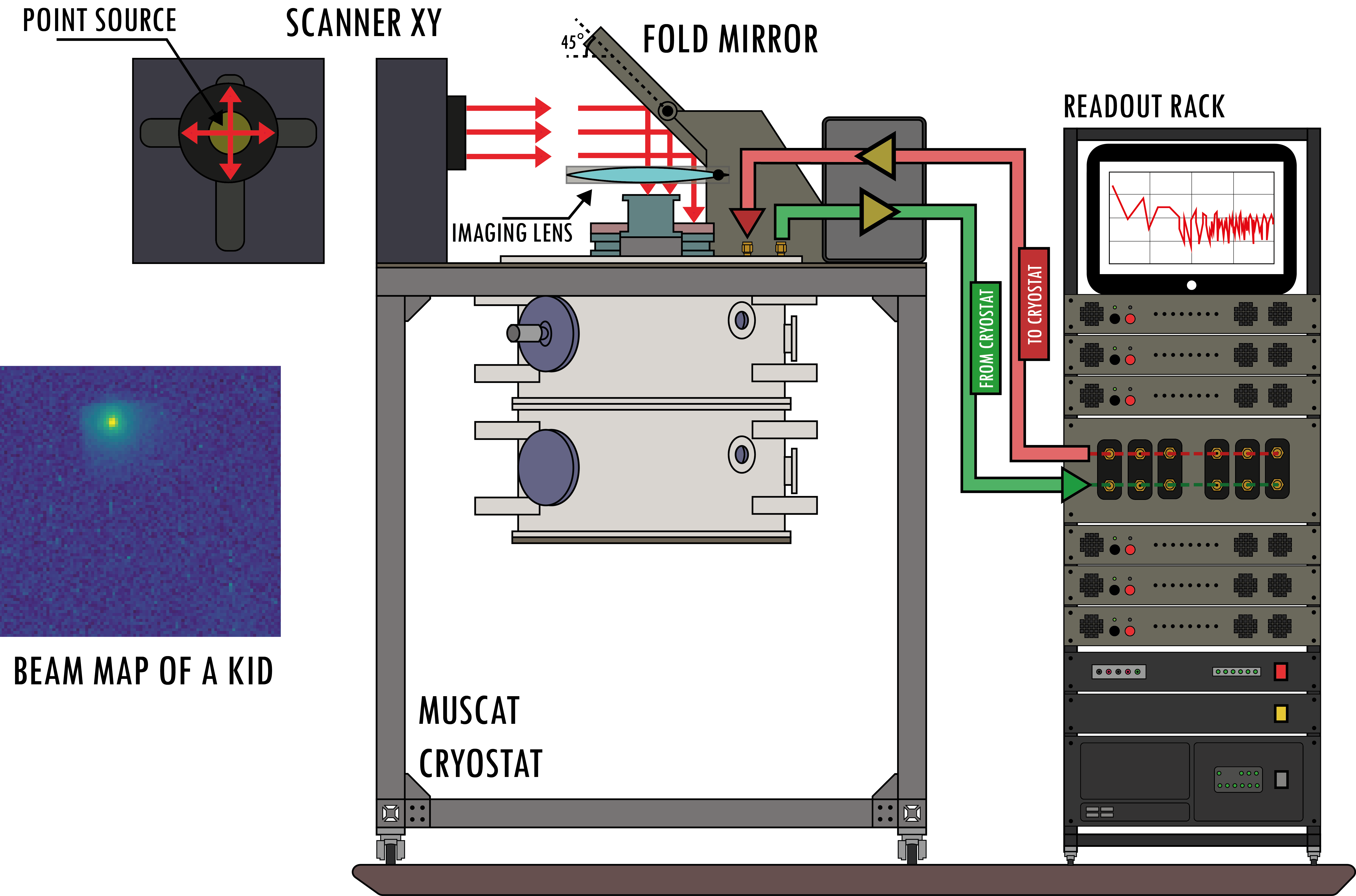}
    \caption{Beam mapping experiment. A fold mirror and image forming lens create an image of the focal plane outside of the cryostat that is scanned using a blackbody source. The response of each detector is measured through the scan providing a beam map of each detector in the arrays. By continuously capturing the response of each resonator and comparing it with the position of the source, the position-intensity maps (left) are generated, that characterise the beam shape of each detector.}
    \label{fig:VNA_meas}
\end{figure}

\subsection{Responsivity and noise measurement}

Intrinsically, two fundamental sensitivity limits are present in a KID detector: the fluctuations in response caused by the generation-recombination of quasiparticles, whose spectra is white up to a cutoff frequency defined by the quasiparticle lifetime $\tau_{\mathrm{qp}}$, and contributions of the two-level system (TLS) noise due to fluctuations of the complex dielectric constant at low frequencies. However, in a complete analysis, other external noise sources must also be taken into account, such as that associated with the cold amplifiers or the warm electronics system.

The standard figure of merit for detector sensitivity is the noise equivalent power (NEP), defined as the signal power that gives an RMS signal to noise of unity in a system that has a bandpass of 1 Hz. The full system NEP is calculated by S/R, where S is the frequency noise spectral density of the measured timestreams and R is the device responsivity (df$_{\theta}$/dP).

The responsivity/noise measurements of the detectors on the MUSCAT focal plane are carried out through the experiment shown in Figure 5. The front facing side of a large copper plate is coated with a mix of carbon-loaded epoxy and 1 mm SiC grains, to obtain high emissivity. It contains embedded thermometers evenly distributed throughout the face. Current-controlled heaters are mounted on the back, enabling stable temperature control over a range of 290--340~K.

\begin{figure}
    \centering
    \includegraphics[width=0.75\textwidth]{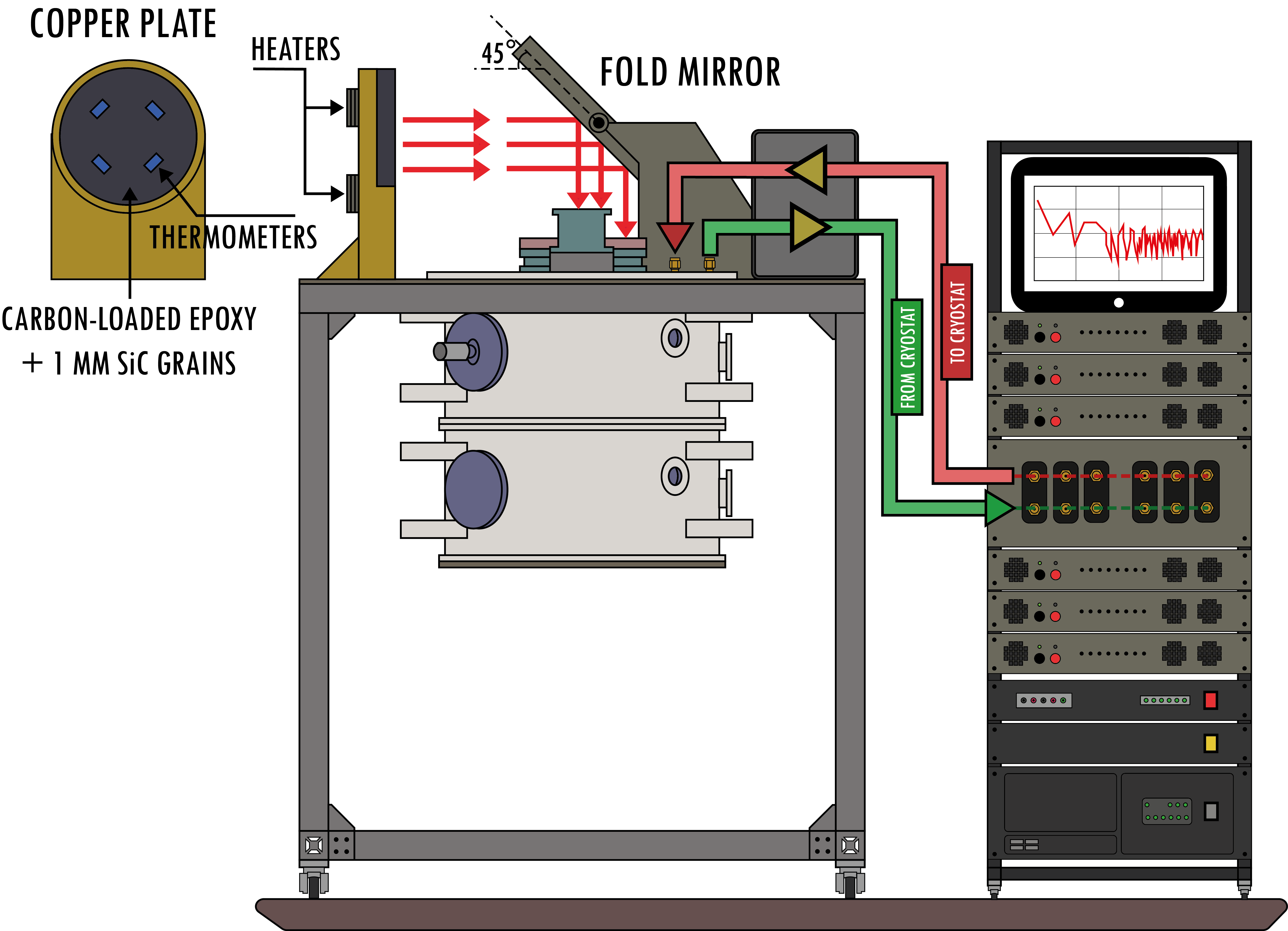}
    \caption{Experiment to measure the detector responsivity/noise. A large copper plate, coated with a mix of carbon-loaded epoxy and 1 mm SiC grains, with thermometers and heaters, acting as a variable source in a range of 290--340~K, is mounted in front of the flat mirror over the cryostat entrance aperture, out of focus. To characterise the responsivity and the detectors NEP, several timestreams are captured for various plate radiation temperatures.}
    \label{fig:VNA_meas}
\end{figure}

Using the same optical system for the beam mapping, we mount the plate at the height of its optical axis, but out of focus, in such a way that, together with its size, it is sufficiently large to fill the beams of every detector fully.

The noise and responsivity estimates are accomplished by measuring the simultaneous response of all detectors to the known radiation of the black-body source, at ten different temperatures between 293~K and 333~K. For each of these temperatures, the readout system is tuned onto the resonances then frequency sweeps and 200-second timestreams are acquired. This is enough time to characterize the contribution of the TLS noise and $1/f$ knee. The sampling frequency of the readout system, 488~Hz, allows characterizing in detail the generation-recombination noise of the quasiparticles, although it is not enough to observe the roll-off and estimate the lifetime $\tau_{\mathrm{qp}}$.

The NEP calculation for the detectors on the focal plane is key to determine their performance and evaluate their ability to detect astronomical sources and their associated integration times (mapping speed).\cite{Brien2020}

\subsection{FTS measurement}

The designed optical bandwidth of MUSCAT is 50 GHz (0.2 mm) centered at 275 GHz (1.1 mm) and is defined by the set of low-pass filters of the instrument Lyot stop at the high end, and by the horn block and cylindrical waveguide that covers the detectors in the focal plane at the low end.\cite{Tom2018}

To characterize the spectral response of the instrument, we performed the experiment described in Figure 6. A moderate-resolution Martin-Puplett Fourier transform spectrometer (FTS) with a high-temperature IR source that operates at 1200~K is mounted on top of the aluminium frame. A fold mirror inclined $45~\si{\degree}$ relative to the horizontal is placed on the opening of the FTS, aligned with the flat mirror on the cryostat aperture used in the previous measurements. While the FTS does not need to be in the focus of the camera, the alignment is adjusted for maximum signal.

By controlling the distance between the mirrors of the FTS, we perform a wide frequency sweep, from 600~MHz to 6~THz with a step of $\sim600~\si{\mega\hertz}$, of the response of a subset of detectors on the focal plane (different channels), continuously acquiring their timestreams. 

\begin{figure}
    \centering
    \includegraphics[width=0.8\textwidth]{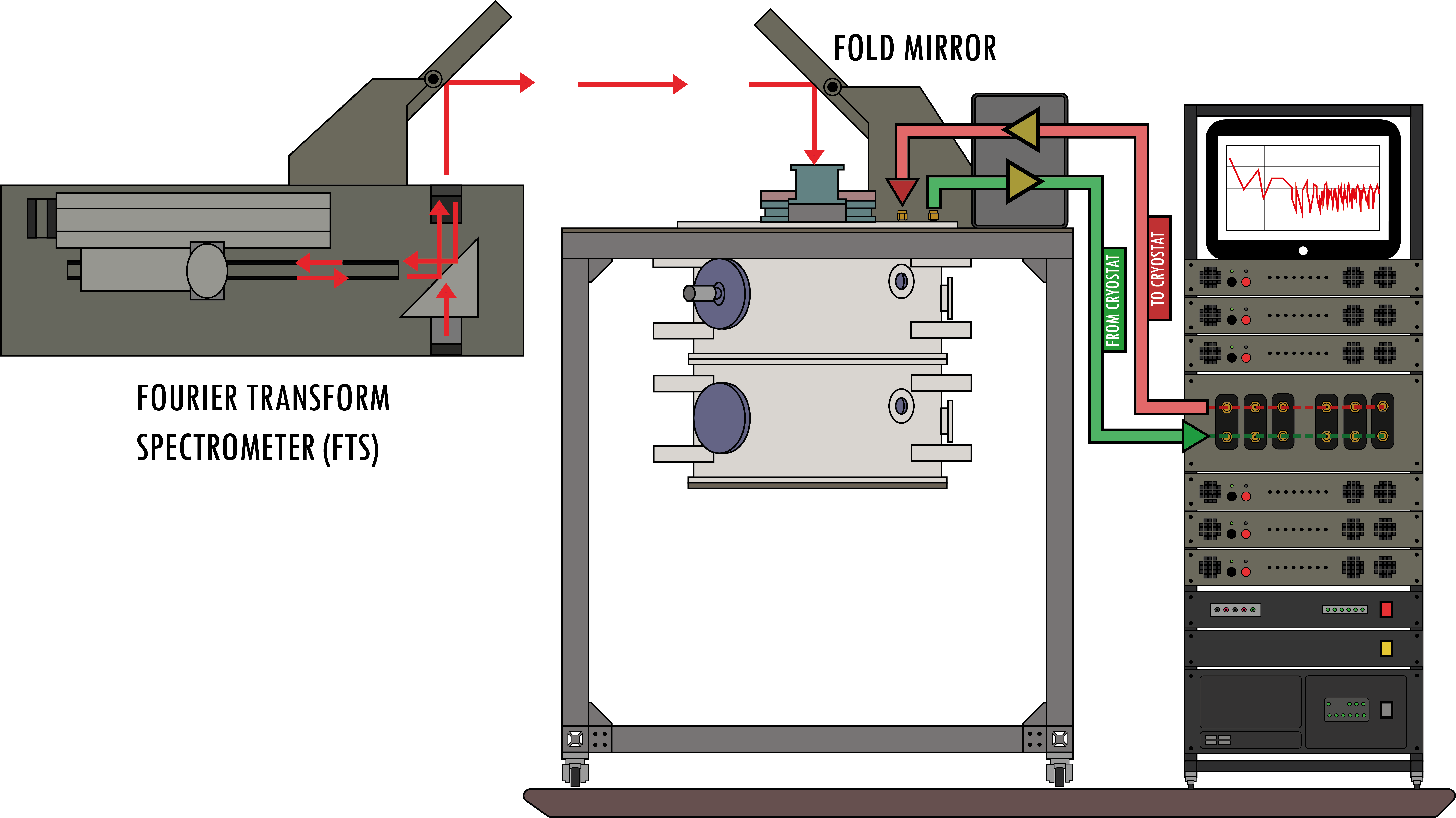}
    \caption{Experiment to measure the instrument spectral response. A Fourier Transform Spectrometer (FTS) optically aligned with the camera, scans the spectral response of the detectors in the range of 0 to 6~THz, with a step of 600~MHz.}
    \label{fig:VNA_meas}
\end{figure}

\section{Focal plane performance results}

\subsection{Responsivity/noise results}

The transmission measurements of the six MUSCAT detector arrays are presented in the graphs of Figure~7, grouped in pairs according to the development to which they belong, channels 1-2 to MD-M, 3-4 to MD-J, and 5-6 an MD-F. The output power of the VNA power was adjusted for each channel to drive the resonators at the desired power (as defined earlier).

\begin{figure}
    \centering
    \includegraphics[width=1\textwidth]{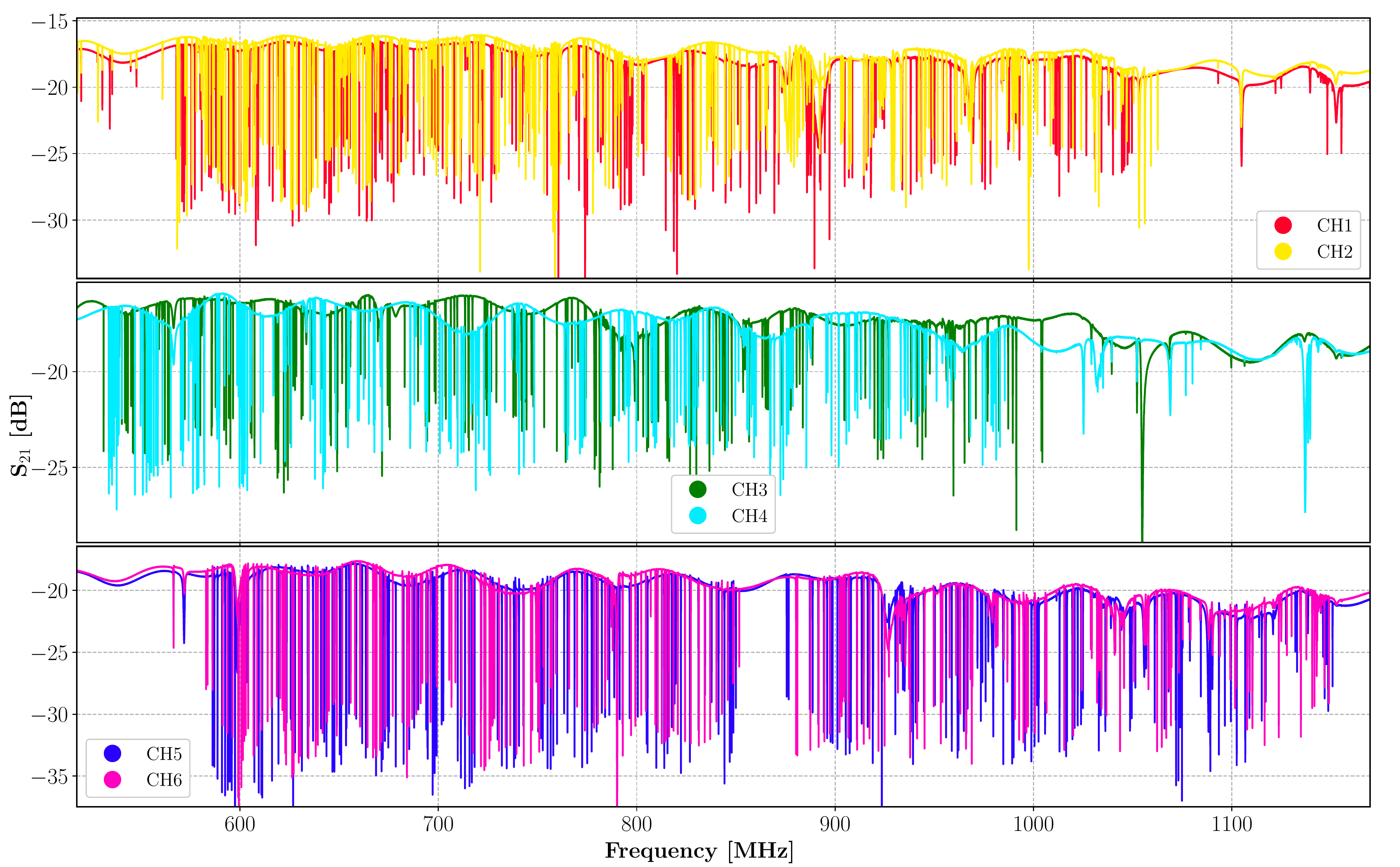}
    \caption{Scattering parameter $S_{21}$ (transmission) of the six MUSCAT channels, grouped according to the fabrication iteration to which they belong: MD-M (upper), MD-J (middle), MD-F (lower).}
    \label{fig:VNA_meas}
\end{figure}

In all the channels we observe consistent baseline levels $\sim$-17~dB, indicative of uniformity in the array fabrication and mounting, the amplifiers, and the input and output coaxial cables and filters inside the cryostat. Additionally, good consistency is observed between channels that belong to the same wafer, showing similar bandwidths and frequency scatter. The resonators of channels 1 and 2 are distributed by a bandwidth of $\sim500~\si{\mega\hertz}$, of $565\mbox{--}1065~\si{\mega\hertz}$; 3 and 4 start at $\sim 530~\si{\mega\hertz}$, channel 3 ends at $1007~\si{\mega\hertz}$ ($\sim$477 MHz of bandwidth), while channel 4 ends at 988~MHz ($\sim458~\si{\mega\hertz}$). Instead, the resonators for channels 5 and 6 are contained between $580\mbox{--}1152~\si{\mega\hertz}$ ($\sim572~\si{\mega\hertz}$). With the effective readout bandwidth of 480~MHz, only channels 3 and 4 can be readout in their entirety, some resonators will be out of band for channels 1 and 2, and a significant amount out for channels 5 and 6. On the other hand, while in channels 5 and 6 the gap between the low and high-frequency bands of the resonators is easily distinguished, in channels 2 and 4 it is reduced, and in the rest it is unrecognizable.

The frequency spread in channels 5 and 6 do considerably exceed the design bandwidth, even though the random scatter in frequencies is lowest in these channels. The rest of the arrays show less overall spread in frequencies, but have more irregular spacing, increasing the number of collisions between resonators. So, most of the resonators of channels 1 to 4 can be read simultaneously.

Close inspection of the sweep data reveals a total of 1974 resonance features. Comparing channels from the same wafer and excluding the least prominent of any resonances that appear at identical frequencies on both channels reduces the total down to 1524: 251 for channel 1, 250 for channel 2, 217 for channel 3, 222 for channel 4, 233 for channel 5, and 219 for channel 6. As mentioned in Section~2.2, it is common that the number of resonators does not match the number of detectors in the array.

Since the required bandwidth of some of the channels exceeds the effective bandwidth of the readout system ($\sim480~\si{\mega\hertz}$), for the complete characterization of the resonators found in each channel, they are divided into two bands of no more $420~\si{\mega\hertz}$ each. For each band, we generate a list of tones and assign it its respective local oscillator frequency.

Under the procedure of Section~2.4, we measured the responsivity and associated noise of all of resonators identified.

The resonators response to changes in the incident radiation power allows us to evaluate whether the resonator is real detector, or if it is a feedline feature. In a KID detector, the increase in radiation power produces a gradual decrease in its resonance frequency, i. e. the slope of the curve f$_{0}$ vs power is negative. In some of the resonators on our tone lists, this response is not observed, or the power increases the resonance frequency, or it is simply chaotic. Therefore, as the first cut filter, we discard those resonators whose slope f$_{0}$ vs power is positive.

On the other hand, we also remove the resonators whose white-noise level is high, with fluctuations in the resonance frequency shift $df$ greater than 1 Hz. In this way, we have 1129 useful resonators distributed among channels 1 to 6 such as 185, 183, 166, 170, 223 and 202, respectively.

We calculated the NEP under a room temperature optical load for these resonators, using the noise spectra obtained under the blackbody radiation at 298.5~K. The results for each channel are presented in the histograms of Figure 8, and are the product of averaging for frequencies $>$10~Hz.

It is noteworthy that the distributions of all the channels peaks at $\sim2.4\times10^{-15}~\mathrm{W\,Hz^{-1/2}}$, and given that the theoretical value of the NEP under a load of $\sim$300~K is $\sim2.7\times10^{-15}~\mathrm{W\,Hz^{-1/2}}$, it shows that the vast majority of MUSCAT detectors are background photon-noise limited. The shape of the distribution points to the same conclusion; in all the channels, from left to right, a sharp rise to the maximum is observed with a smoother descent. This behaviour reveals the low probability of find detectors with noise levels below the maximum because we have run into the fundamental limit. Although Gómez-Rivera et al.\cite{Gomez2020} have already shown that MUSCAT detectors are photon-noise limited under lower loads (10--70~K), our result is of great relevance because, in addition to confirming this, it shows that the noise contributed by other elements of the system (amplifiers/readout system) is negligible.

\begin{figure}
    \centering
    \includegraphics[width=1\textwidth]{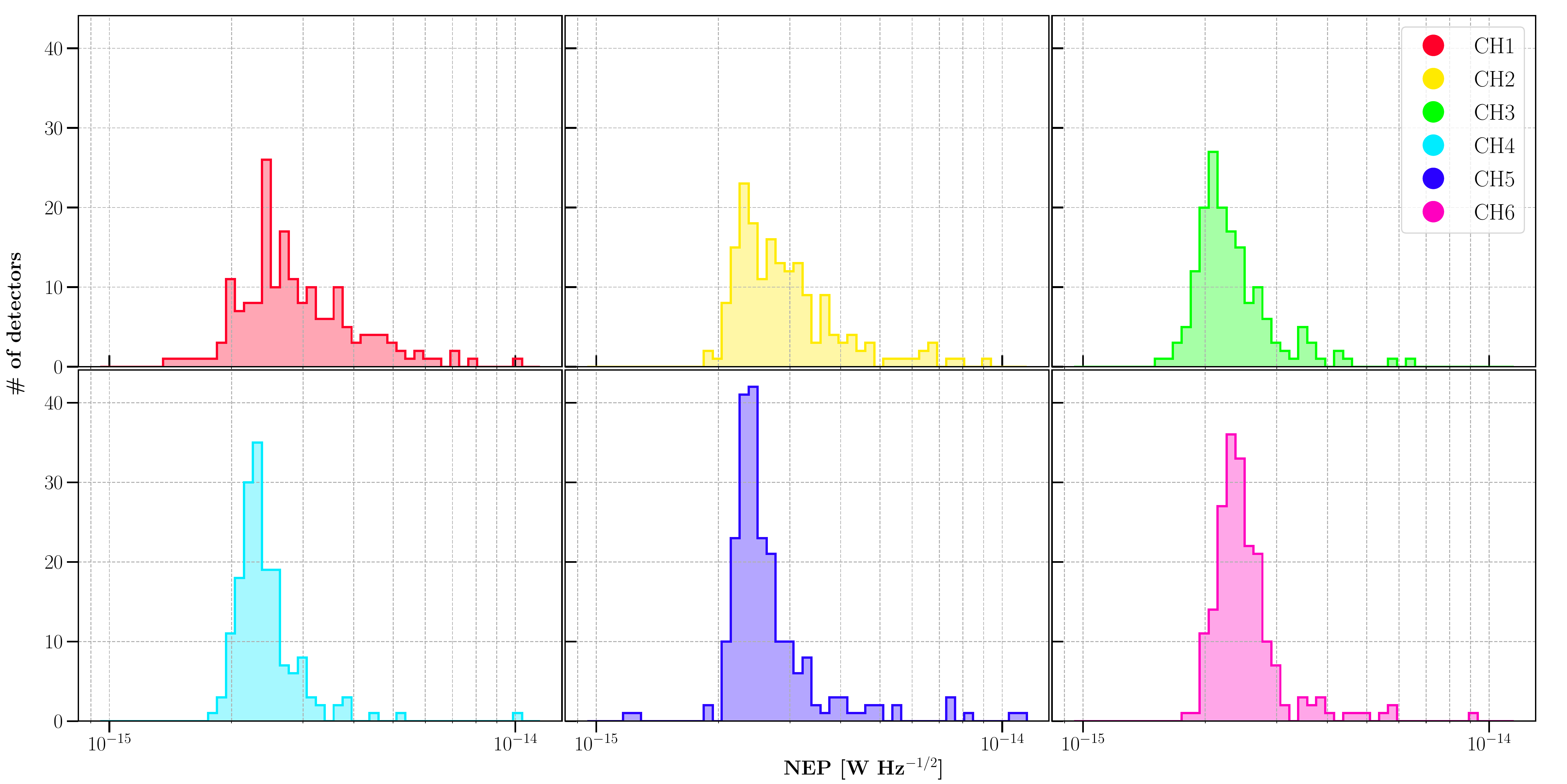}
    \caption{Histograms of the NEP measured under an optical load of $\sim$300 ~ K for each of the MUSCAT channels. The maximum accumulation of the distributions at $\sim2.4\times10^{-15}~\mathrm{W\,Hz^{-1/2}}$, very similar to the estimated theoretical noise of $\sim2.7\times10^{-15}~\mathrm{W\,Hz^{-1/2}}$, and the sharp descent of the curves towards less noise, show that the vast majority of the detectors in the MUSCAT focal plane are photon-noise limited.}
    \label{fig:VNA_meas}
\end{figure}

Through long exposure timestreams, it is also possible to characterize the frequency from which the 1/f noise becomes dominant (1/f knee). For this, a PCA (principal component analysis) filter is applied to them, their spectrum is obtained, and the intersection with white noise (generation-recombination noise) is calculated. In the histogram of Figure~9, we present the accumulated $1/f$ knee values for each channel.

The total distribution abruptly peaks at 0.1~Hz with a smoothed decline. Instead, the distributions of each channel have their own behaviour, shared only between arrays of the same wafer. A trend is also observed in the increase of the $1/f$ knee with the channel number, thus channel 1 peaks at 0.08~Hz, channel 2 at 0.14~Hz (the midpoint of a plateau of maximum values), channel 3 a 0.1~Hz, channel 4 at 0.14~Hz, and for channels 5 and 6 at 0.27~Hz and 0.55~Hz respectively.

\begin{figure}
    \centering
    \includegraphics[width=1\textwidth]{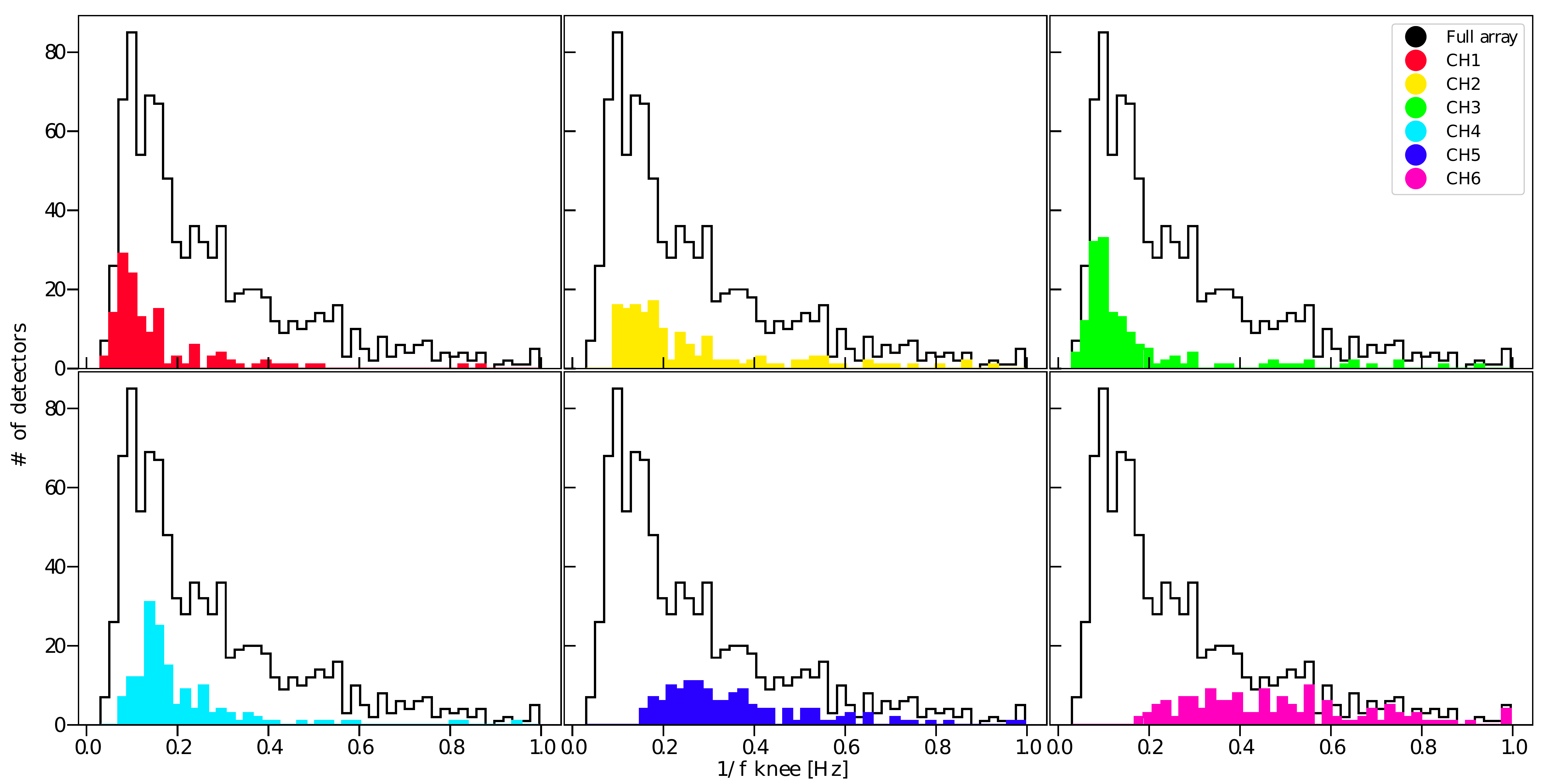}
    \caption{Histograms of the 1/f knee of each MUSCAT channel. The total distribution (solid black line) peaks at 0.1 Hz, although the contributions of each channel move towards a higher frequency reaching maximums from 0.07 Hz (channel 1) to 0.55 Hz (channel 6).}
    \label{fig:VNA_meas}
\end{figure}

\subsection{Beam shape results}

Following the procedure described in Section~2.3, we obtain the beam maps of the resonators. These maps represent the response of the resonator to a point source as a function of its position in the focal plane. We fit a Gaussian 2D function to each of them, obtaining the position and the characteristic FWHM width of the beam of each resonator. This allows us to link the response of each resonator with the physical position of a detector in the array, as well as to identify with absolute certainty false resonators due to cross-talk and feedline features that the criteria of the previous selection do not cover. Thus, of the total of resonators loaded into the reading system, only a fraction belongs to real detectors.

We first discard the beam maps of those whose background noise significantly modifies the beam shape. For this, we measure the signal to noise in each map and fit a 2D elliptic Gaussian function, where we extract the widths of the major/minor axes and calculate the associated eccentricity ($e$). The signal-to-noise varies according to the array number, $\sim$500 for channels 5 and 6, $\sim$200 for channels 1 and 2; hence we define a highly noisy detector as one whose SNR$<$15. On the other hand, since the shape of the design beam is circular, we consider that its shape is moderately altered if $e>$0.6 and highly if $e>$0.75. Therefore, taking both criteria, we discard all detectors that meet the condition ($e>$0.75)$\cup$($e>$0.6 $\cap$SNR$<$15).

In Figure 10, we show the FWHM histograms of the beams associated with these resulting detectors, indicating the contribution of each of the channels to the total. The most notable characteristic in the distribution is the narrowing and reduction of the dispersion as we move from channel 1 to 6, which also causes a slight asymmetry in the total distribution. This reveals greater homogeneity in the beam shape for the central (3 and 4) and left lateral (5 and 6) channels. Following this behaviour, we discard the end detectors: those with a width less than 5$\sigma$ of the distribution of channels 5 and 6, and greater than 3$\sigma$ of the distribution of channels 1 and 2 with SNR$<$35.

On the other hand, the spacing criterion between resonators is not sufficient to rule out probable collisions. To complement the analysis, we study the beam maps one by one, discarding those with two or more spots in it.

\begin{figure}
    \centering
    \includegraphics[width=1\textwidth]{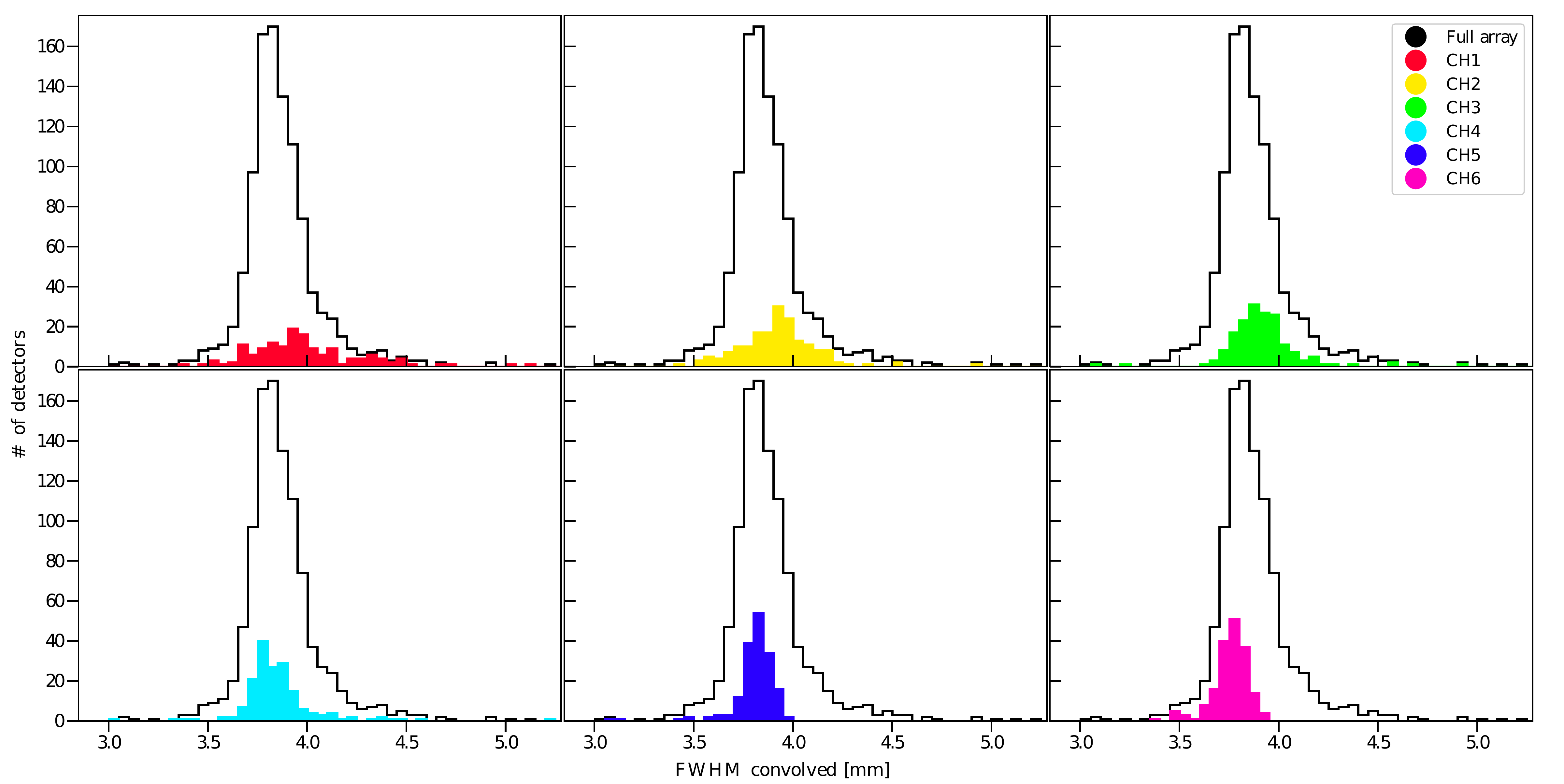}
    \caption{Histogram of the FWHM of the detector beams convolved with the source aperture (2.7 mm) of all the arrays (black solid line), in comparison with the histograms per channel (colours). Greater dispersion is observed in the distribution, accompanied by the widening of the beam, as we move from channel 6 to 1.}
    \label{fig:VNA_meas}
\end{figure}

On the other hand, the collision criterion between resonators applied in the previous section is not sufficient to identify all interactions between detectors. To complement the analysis, we study the beam maps one by one in search of two or more spots on the same map.

For the resulting detectors (cleaned detectors), we deconvolved the beam maps with the source aperture (2.7~mm in diameter), and we calculated the FWHM (2D Gaussian fit) and the eccentricity (2D Elliptical Gaussian) of the characteristic beam of each cleaned detector. In the histograms in Figure 11, we present by channel, the deconvolved FWHM distribution and the associated eccentricity, as well as the centred and stacked beam maps of the arrays. As we have already observed previously (Figure~10), there is a trend for the beam to narrow towards channels 5 and 6, but now with a slightly higher dispersion, since the wider beams are less susceptible to convolution with the aperture. The distribution peaks at FWHM$\sim$3.27~mm, slightly above the design value of 3.08~mm ($F\lambda=2.8\times1.1~\mathrm{mm}$). Regarding shape, the eccentricity peaks at $\sim$0.5, that is, a typical pixel has a subtle oval shape, with a major/minor axis ratio of 1.15.

\begin{figure}
    \centering
    \includegraphics[width=1\textwidth]{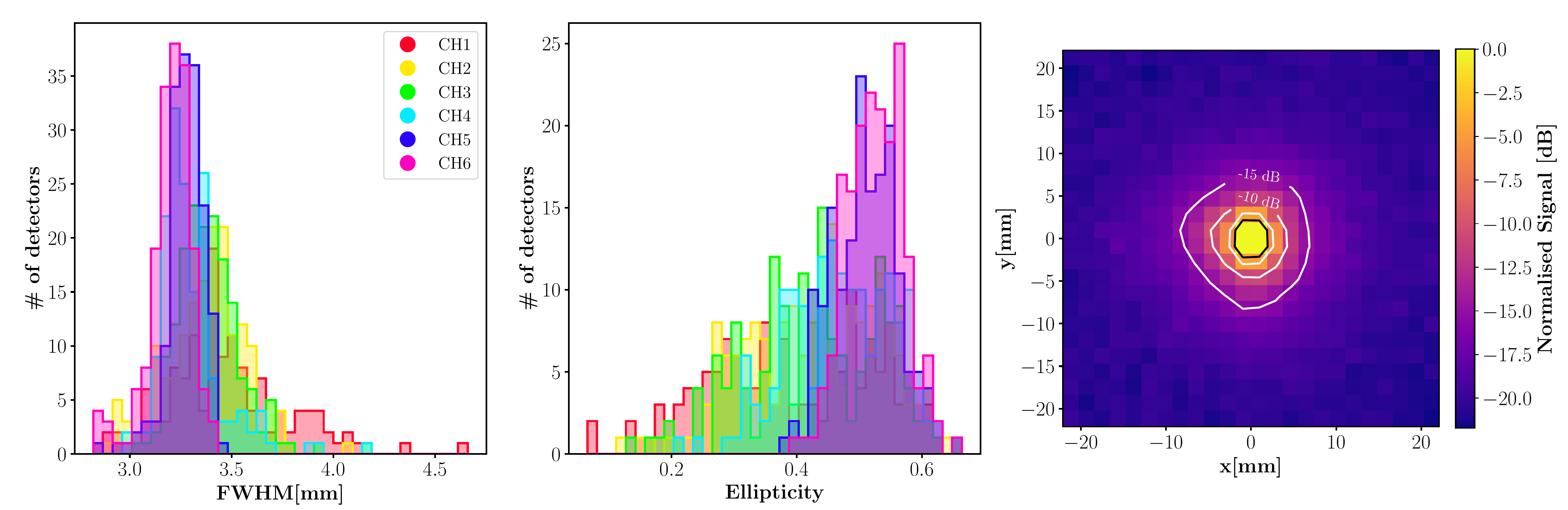}
    \caption{Histograms of the FWHM (left) and the eccentricity (middle) of the deconvolved beam shape. The beams of the detectors tend to widen and have a more circular shape (less eccentricity) as we move from channel 6 to 1. Normalized centred and stacked beam shape of MUSCAT detectors (right). The typical MUSCAT pixel has a 3.27mm FWHM (solid black line), with a slightly oval shape ($e\sim$0.5).}
    \label{fig:VNA_meas}
\end{figure}

On the other hand, in Figures~12 and 13 we present the results of the fitted beam shapes of each detector on the focal plane as heat maps of the FWHM (2D Gaussian fit) and its eccentricity (2D Elliptical Gaussian fit), respectively. The most notable feature in both graphs is the reduction in elongation followed by the widening of the beam shape towards the top of channel 1 (upper left in the graph). This correlation along a well-defined direction suggests a slight tilt or deformation of the focal plane relative the image plane in that region, blurring its detectors, and causing the characteristic spot of the focal plane that we observe in both graphs. It is not clear if this focal plane distortion comes from the lab or cold optics, it will be necessary to observe its behaviour in the telescope but should be possible to correct for in the data pipeline.

This "spot" has an almost zero effect towards channels 5 and 6 (left); hence their beams are much more homogeneous (less dispersion); In contrast, in channels 1 and 2 (right) there are two populations delimited by this spot, one wide and circular inside, and the other narrow and elongated outside, i. e. there is a greater variety of beam sizes (greater dispersion).

\begin{figure}
    \centering
    \includegraphics[width=1\textwidth]{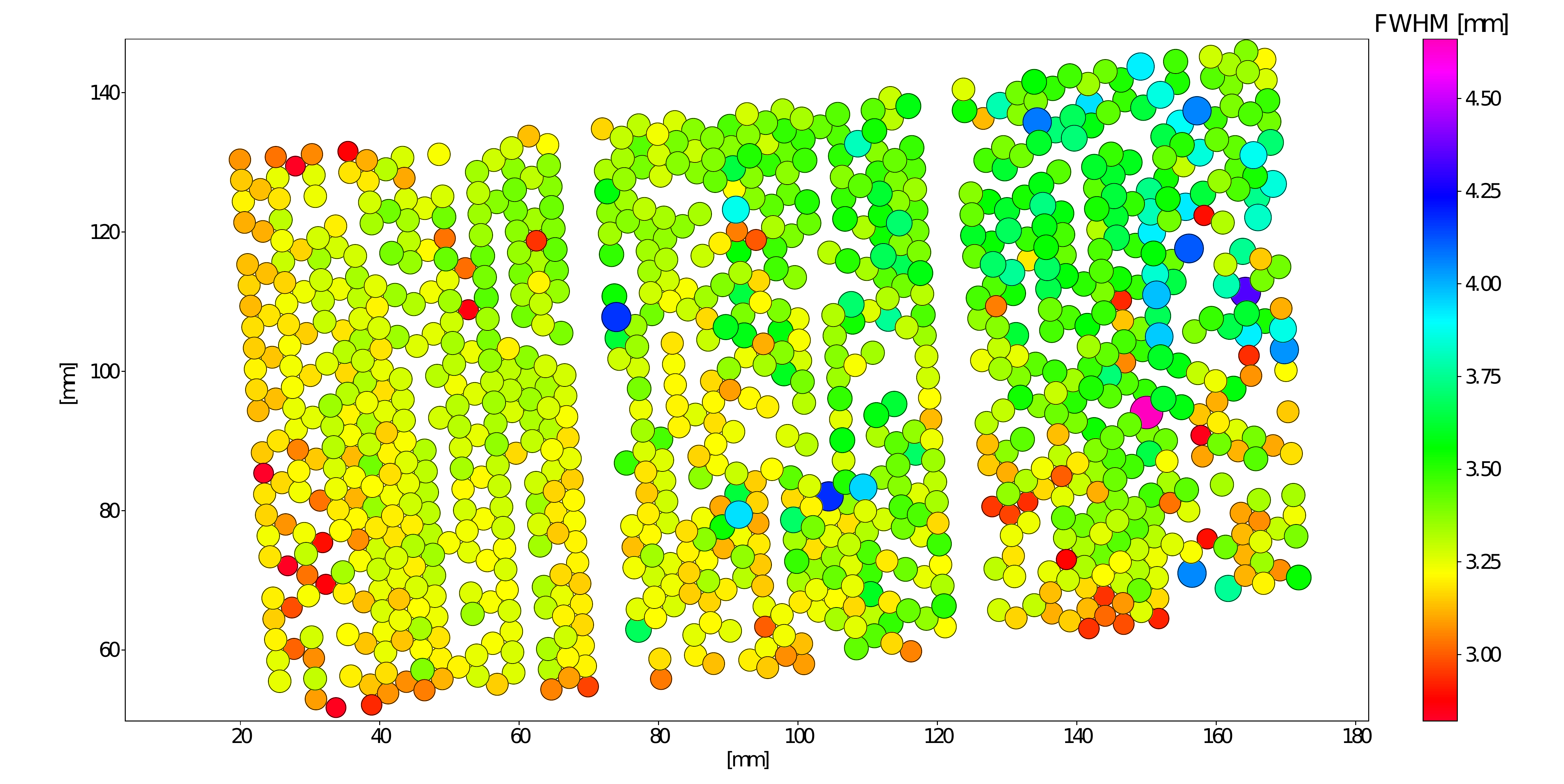}
    \caption{FWHM deconvolved heat map of MUSCAT focal plane detectors (channels 6 to 1 from left to right). A clear trend is observed in the widening of the beam as we approach the upper part of channel 1 (upper right corner). The diameter of each pixel equals its fitted FWHM from the 2D Gaussian function.}
    \label{fig:VNA_meas}
\end{figure}

\begin{figure}
    \centering
    \includegraphics[width=1\textwidth]{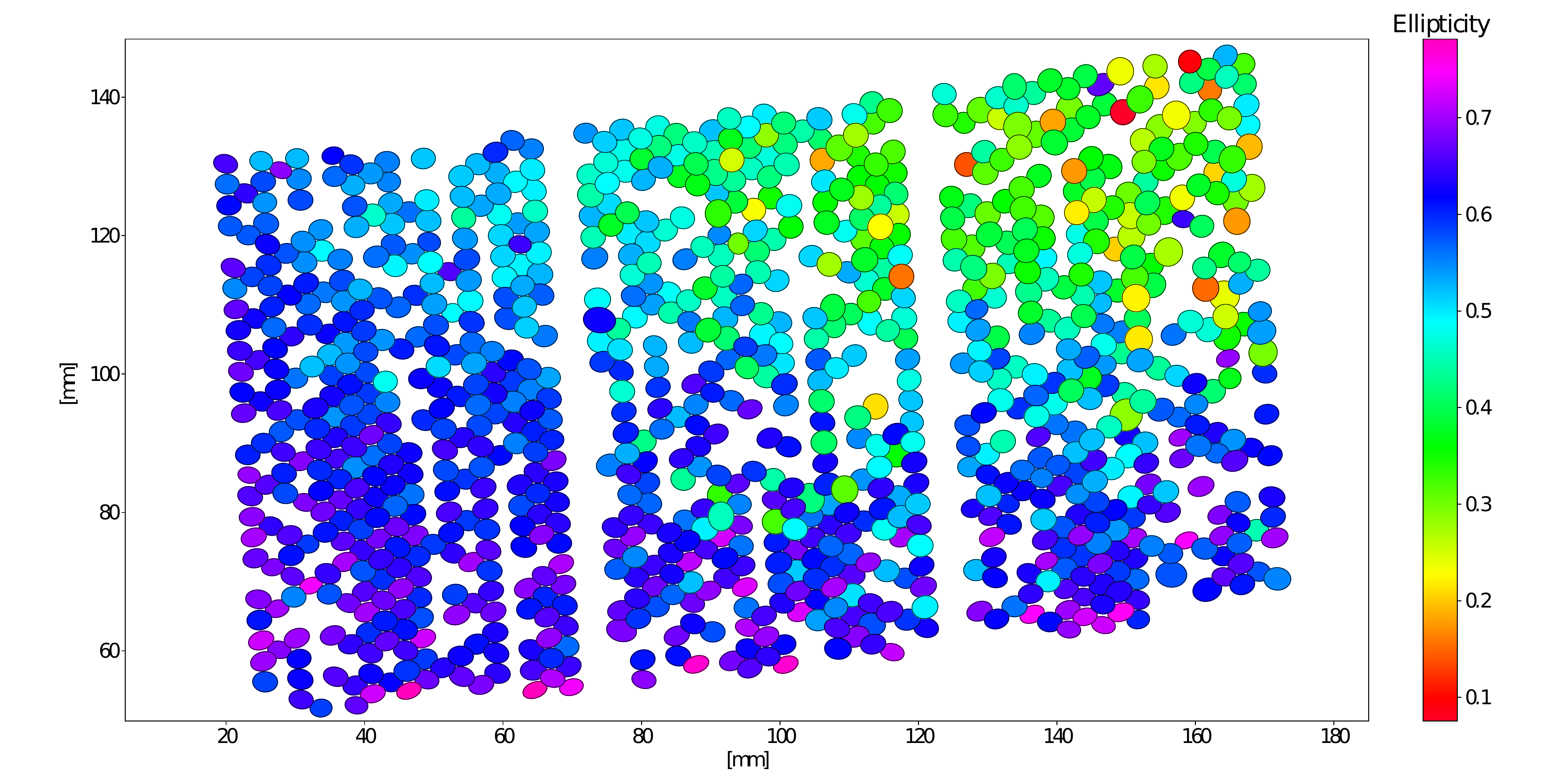}
    \caption{Heat map of the eccentricity of the MUSCAT focal plane detectors (channels 6 to 1 from left to right). A clear trend is observed in the beam to acquire a more circular shape as we approach the upper part of channel 1 (upper right corner), accompanied by a greater widening. The dimensions of each pixel are equal to the FWHM of the major/minor axes of the fitted beam shape, from the 2D Gaussian Elliptical function.}
    \label{fig:VNA_meas}
\end{figure}

\subsection{Spectral response}
\label{sec:fts}

The measured and modelled optical response of the MUSCAT detectors are presented in Figure 14. Detectors from all six readout channels, across three separate fabrication runs all show exactly the same cut-on and cut-off frequencies. The detector responses are very sensitive to any misalignment of the FTS, which is the likely cause of the variations across the centre of the band.  The modelled data on the right of Figure 14 consists of the measured transmission profiles of the optical filters (measured warm) combined with the simulated performance of the horn block from HFSS. The measured spectral response has been normalized by constant area against modelled spectrum and corrected by the profile of the black-body source used (1200~K).

The agreement between the measured and expected spectrum is good. The measurements show a well defined band of $\sim$50~GHz, from 250~GHz (1.2~mm) to 300~GHz (1mm), with a slight shift to the lower frequency relative to the model. The abrupt  waveguide cutoff at the low end, and the less abrupt cut off at the high end due to the filter stack, are clearly visible.

\begin{figure}
    \centering
    \includegraphics[width=1.0\textwidth]{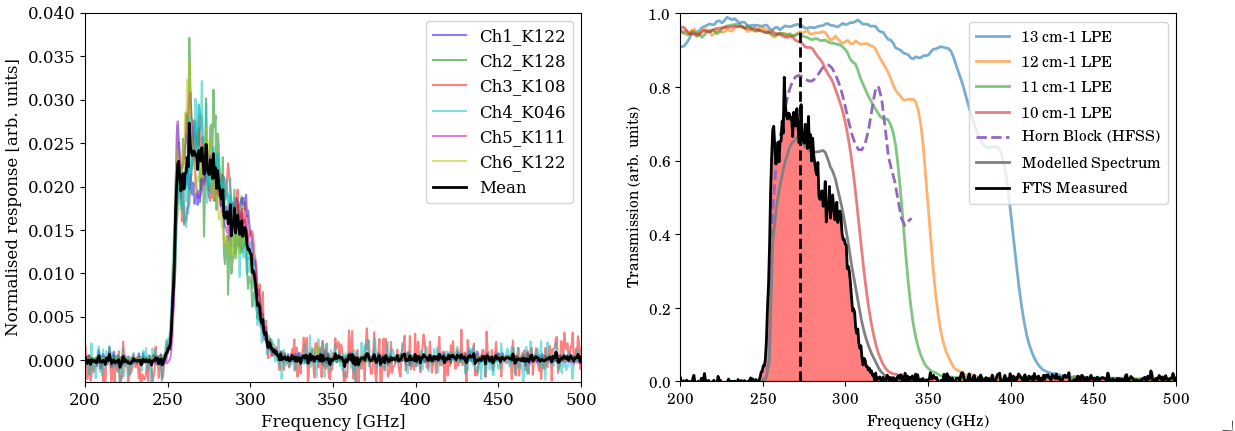}
    \caption{(Left) The FTS measured optical response of six detectors, one from each of the MUSCAT readout channels, and their mean. (Right) A comparison of a single measured spectrum with a combination of measured optical filter bandpasses and HFSS simulated detector response. The measured spectrum is normalised to match the area of the modelled spectrum and the shapes are in good agreement.}
    \label{fig:FTS}
\end{figure}

\subsection{Summary of yield}
 Microwave transmission, sensitivity, beam profiles and optical spectra have been measured across the MUSCAT focal plane. An initial resonator search identified 1974 features and after exclusion of 450 artefacts from cross-talk between readout channels, 396 low sensitivity devices and 86 devices with highly distorted beam shape, an 81 devices that are are outside of the readout electronics bandwidth, we ascertain that 961 out of the 1458 physical detectors are suitable for operation. Table 1 details the yields per channel. These numbers do not represent the final yield of the MUSCAT instrument as the sky load at the LMT site is quite a bit lower than that of the 300K lab. The procedures outlined above will have to be repeated during the commissioning of MUSCAT.

\begin{table}[htbp]
	\centering
	\begin{center}
		\begin{tabular}{|c|c|c|c|c|c|c|c|}
		\hline

		\textbf{Description} & 
		\textbf{Ch.~1} & 
		\textbf{Ch.~2} & 
		\textbf{Ch.~3} & 
		\textbf{Ch.~4} & 
		\textbf{Ch.~5} & 
		\textbf{Ch.~6} &
		\textbf{Total} \\
		\hline

		Physical detectors & 243 & 243 & 243 & 243 & 243 & 243 & 1458 \\ \hline
		Initial search & 432 & 427 & 261 & 275 & 297 & 282 & 1974 \\ \hline
		Filter cross-talk & 283 & 277 & 233 & 254 & 252 & 225 & 1524 \\ \hline
		Filter sensitivity & 185 & 183 & 166 & 170 & 223 & 202 & 1129 \\ \hline
		Filter beam quality & 151 & 154 & 160 & 163 & 216 & 198 & 1042 \\\hline
	    \begin{tabular}[c]{@{}l@{}}Filter readout bandwidth \\(Total usable detectors) \end{tabular} & 146 & 150 & 157 & 163 & 182 & 163 & 961 \\\hline
		
		\end{tabular}
		\caption{Detector array yield per channel in laboratory setting.}
		\label{tabla:sencilla}
	\end{center}
\end{table}

\section{Conclusions}
\label{sec:conclusions}

Through MUSCAT's focal plane characterization we have shown that most of their detectors, $\sim$1000, operating continuously at a base temperature of 130~mK, are photon-noise limited under an optical load of $\sim$300~K, in turn validating the operation of the cold electronics of the cryostat and the room temperature readout system. The 1/f noise from the detectors only becomes dominant below 0.1 Hz. The pixels are almost circular in shape with a typical width of 3.27 mm, just above the expected value. On the other hand, the good agreement of the spectral response of the detectors concerning the design was also verified, covering a bandwidth of 50~GHz (2 mm) centred on $\sim$275~GHz (1.1 mm). With this validation of the focal plane of the instrument, we show that MUSCAT is ready for the installation and operation at the LMT.

\acknowledgments   
 
We acknowledge RCUK and CONACYT through the Newton Fund project ST/P002803/1, CONACYT for  support the fellowship for the instrument scientist (grant no. 053), Chase Research Cryogenics for the development of the sub-kelvin coolers and to XILINX Inc. for the donation of the FPGAs used for the ROACH2 boards.

\end{document}